\documentclass[aps,reprint,nofootinbib,nobibnotes,notitlepage,superscriptaddress,twocolumn,prl,
 amsmath,amssymb
]{revtex4-2}

\usepackage[caption=false]{subfig}
\usepackage{braket}
\usepackage{float}
\usepackage{lipsum}
\usepackage{graphicx}
\usepackage{dcolumn}
\usepackage{bm}
\usepackage{natbib}
\usepackage{hyperref}
\hypersetup{
	colorlinks = true,
    linkcolor = Red,
    urlcolor  = Red,
    citecolor = Red
}

\usepackage{microtype}

\usepackage{verbatim}
\usepackage[amssymb]{SIunits}
\usepackage{tabularx}
\usepackage[dvipsnames]{xcolor}
\usepackage{wasysym}

\usepackage{tikz}
\usepackage{soul}

\def\be{\begin{equation}}
\def\ee{\end{equation}}

\usepackage{color}

\definecolor{darkgreen}{RGB}{0,120,0}

\newcommand{\resub}[1]{{#1}}
\newcommand{\Mpch}{h^{-1}\mathrm{Mpc}}

\newcommand{\av}[1]{\left\langle{#1}\right\rangle}

\newcommand{\vl}{\vec l}

\newcommand{\hn}{\hat{\vec n}}

\newcommand{\vtheta}{\boldsymbol{\theta}}

\def\beq{\begin{eqnarray}}
\def\eeq{\end{eqnarray}}
\let\vec\mathbf
\usepackage{xspace}

\usepackage{empheq}

\newcommand{\es}[2] {\begin{equation} \label{#1} \begin{split} #2 \end{split} \end{equation}}

\def\aap{\rm{A\&A}}

\def\apj{\rm{ApJ}}                 

\def\apss{\rm{ApSS}}

\begin{document}

\title{Too Hot to Handle: Searching for Inflationary Particle Production in \textit{Planck} Data}

\author{Oliver~H.\,E.~Philcox}
\email{ohep2@cantab.ac.uk}
\affiliation{Simons Society of Fellows, Simons Foundation, New York, NY 10010, USA}
\affiliation{Department of Physics, Columbia University, New York, NY 10027, USA}
\affiliation{Department of Physics,
Stanford University, Stanford, CA 94305, USA}

\author{Soubhik Kumar}
\email{soubhik.kumar@nyu.edu}
\affiliation{Center for Cosmology and Particle Physics, Department of Physics, New York University, New
York, NY 10003, USA}

\author{J.~Colin Hill}
\email{jch2200@columbia.edu}
\affiliation{Department of Physics, Columbia University, New York, NY 10027, USA}

\begin{abstract} 
    \noindent Non-adiabatic production of massive particles is a generic feature of many inflationary mechanisms. If sufficiently massive, these particles can leave features in the cosmic microwave background (CMB) that are not well-captured by traditional correlation function analyses. We consider a scenario in which particle production occurs only in a narrow time-interval during inflation, eventually leading to CMB hot- or coldspots with characteristic shapes and sizes. Searching for such features in CMB data is analogous to searching for late-Universe hot- or coldspots, such as those due to the thermal Sunyaev-Zel'dovich (tSZ) effect. Exploiting this data-analysis parallel, we perform a search for particle-production hotspots in the \textit{Planck} PR4 temperature dataset, which we implement via a matched-filter analysis. Our pipeline is validated on synthetic observations and found to yield unbiased constraints on sufficiently large hotspots across $\approx 60\%$ of the sky. After removing point sources and tSZ clusters, we find no evidence for new physics and place novel bounds on the coupling between the inflaton and massive particles. These bounds are strongest for larger hotspots, produced early in inflation, whilst sensitivity to smaller hotspots is limited by noise and beam effects. Through such methods we can constrain particles with masses $\mathcal{O}(100)$ times larger than the inflationary Hubble scale, which represents possibly the highest energies ever directly probed with observational data.
\end{abstract}

\maketitle

\section{Introduction}
\noindent The inflationary paradigm is a leading candidate for explaining the origin of the primordial density fluctuations that eventually seed the anisotropies and inhomogeneities in the cosmic microwave background (CMB) and large-scale structure (LSS), respectively. Cosmological observations are so far consistent with an almost scale-invariant, Gaussian, and adiabatic spectrum of primordial fluctuations, as predicted by some of the simplest models of inflation. Whilst microphysical models of inflation are aplenty~\cite{Martin:2013tda}, we are far from identifying the fundamental physics of inflation (for a recent review on related aspects, see~\cite{Achucarro:2022qrl}). To make progress, a promising and somewhat model-agnostic approach is to identify certain broad-brush features of microscopic theories and understand their implications in the context of current and upcoming cosmological surveys.

Inflationary particle production is one such phenomenon that is present in a wide variety of multi-field scenarios, but has received less data analysis focus thus far than, {\it e.g.}, primordial non-Gaussianity or primordial gravitational waves. The existence of multiple fields during inflation, which may not all be dynamical, is a generic possibility commonly predicted in UV models of particle physics, as well as string theory. A multi-field potential energy landscape involving all these fields can take various forms. Importantly, there are key observational differences that allow us to distinguish among different landscapes depending on the masses of fields. For example, light fields with masses much smaller than the inflationary Hubble scale $H_I$ can contribute to both curvature and isocurvature fluctuations. Particles with masses of order $H_I$ are produced less frequently but can give striking oscillatory signatures; this has been the focus of the recent ``cosmological collider'' program~\cite{Chen:2009zp, Arkani-Hamed:2015bza,Philcox4pt1}. On the other hand, extremely heavy fields are typically not excited, remaining stuck at the bottom of their potential, and can be ``integrated out'' from the inflationary dynamics.

Interestingly, the distinction amongst these possibilities is not rigid. As the inflaton field moves around in the landscape, the (effective) masses of different fields can change. For example, there could be instances when some fields become lighter than usual; in this case, the fields can be produced during such epochs, with particle production shutting off as the masses of the fields increase again. This possibility, especially when massive particles become massless for a finite time during inflation, has been studied extensively in the literature~\cite{Chung:1999ve, Kofman:2004yc, Romano:2008rr, Barnaby:2009mc, Green:2009ds, Barnaby:2009dd, Chantavat:2010vt, Pearce:2016qtn} and typically leads to a bump in the curvature perturbation power spectrum.
On the contrary, the scenario when the massive particles become lighter, but not massless, has been studied less extensively~\cite{Kofman:2004yc, Mirbabayi:2014jqa, Flauger:2016idt}.
In this case, the production of heavy particles is rarer if their minimum mass is still much larger than $H_I$.
However, owing to their large time-dependent masses, the heavy particles can modify the gravitational potential around their locations~\cite{Fialkov:2009xm, Maldacena:2015bha}, which eventually gives rise to localized hot or cold spots in the CMB. The properties, such as the number, temperature profiles, and distribution of these spots have been studied in detail in~\cite{Kim:2021ida, Kim:2023wuk}, a summary of which is given in the next section.

In~\cite{Kim:2021ida, Kim:2023wuk} different strategies for searching for such localized spots were also explored, relying on simulated CMB maps. Several position-space methods were studied, including a matched-filter analysis, which involved searching for such spots directly in the simulated sky maps. It was found that these map-level methods could provide complementary sensitivity compared to searches via the two- or higher-point momentum-space correlation functions. In this context, we emphasize that~\cite{Munchmeyer:2019wlh} performed an $N$-point correlation function analysis using \emph{WMAP} data for scenarios with periodic particle production.
There it was shown that such an $N$-point correlation function analysis is equivalent to a profile-finding (\textit{i.e.}, matched-filter analysis) when profiles do not overlap.
In contrast, in~\cite{Kim:2021ida, Kim:2023wuk} and the present work, we focus on a single instant of particle production when the CMB-observable modes exit the horizon, as well as allowing for scenarios where overlapping profiles are present.

More specifically, we perform a matched-filter analysis using the {\it Planck} PR4 data~\cite{Planck:2020olo}.
We utilize the fact that searches for localized hot- or coldspots share analogies with existing searches for galaxy clusters in CMB maps via the thermal Sunyaev-Zel'dovich (tSZ) effect, {\it i.e.}, inverse-Compton scattering of CMB photons off hot electrons in the intracluster medium~\cite{SZ1969}.\footnote{State-of-the-art current tSZ cluster catalogs include Refs.~\cite{ACT:2020lcv,SPT:2023tib,Planck:2015koh}.}  In both cases, an angular profile for the signal is specified: in the hotspot case, the profile is computed from first-principles inflationary calculations~\cite{Kim:2021ida, Kim:2023wuk}, whilst in the tSZ cluster case, the profile is modeled using functional forms calibrated by deep X-ray or tSZ observations of individual clusters~\cite{Arnaud2010,Planck2013PPP,Battaglia2012}.  Given the theoretical profile shape and knowledge of the data covariance, one can then construct and apply a matched filter~\cite{Haehnelt:1995dg} to the CMB maps to search for features with this profile.\footnote{Analogous methods hold for point source detection, where the relevant profile is simply the instrument's beam profile.}  A key difference between the hotspot and tSZ analyses, however, is that the inflationary hotspots possess the same blackbody spectral energy distribution (SED) as the (other) primary CMB fluctuations, whilst the tSZ effect produces a characteristic non-blackbody spectral distortion.  One can thus exploit not only the angular profile of the signal, but also its SED, in such searches, via the use of multi-frequency matched-filter (MMF) techniques~\cite{Herranz:2002kg,Melin:2006qq}; \resub{this additionally allows the hotspots and thermal SZ clusters to be practically distinguished}.  Below, we adapt existing MMF tools that have been used for tSZ cluster-finding in \emph{Planck} data to search for inflationary particle production hotspots, taking advantage of these analysis similarities.  

In the remainder of this work, we first discuss the particle production hotspots and their phenomenology before presenting our pipeline for their detection. After validating the pipeline on realistic simulations (containing both single and pairwise hotspots), we then apply it to \textit{Planck} PR4 data, and discuss future prospects. Throughout, we adopt the fiducial cosmology based on \citep{Planck:2018vyg}: $\{H_0=67.32 \,\, {\rm km/s/Mpc}, \omega_{\rm b}=0.022383, \omega_{\rm cdm}=0.12011,\tau_{\rm reio}=0.0543,\sum m_\nu=0.06 \,\, {\rm eV} \}$. 
We use the metric convention $\eta_{\mu\nu} = {\rm diag}(-,+,+,+)$.

\section{Hotspots from Particle Production}
\noindent Let us consider a field $\sigma$, whose mass $m_\sigma$ varies in time, which itself is parametrized by the (homogeneous) inflaton $\phi$. Expanded around the minimum, which occurs at $\phi_\star$, we can write
\es{eq:eff_mass}{
m_\sigma(\phi) \approx m_\sigma(\phi_\star) + \frac{1}{2}m_\sigma^{\prime \prime}(\phi_\star)(\phi-\phi_\star)^2 + \cdots,
}
where primes denote derivatives with respect to $\phi$. This implies the Lagrangian containing the massive field $\sigma$ can be written as~\cite{Kim:2021ida,Kim:2023wuk},
\es{eq:lag}{
{\cal L}_\sigma = -\frac{1}{2}(\partial_\mu \sigma)^2 - \frac{1}{2}\left((g\phi-\mu)^2+M_0^2\right)\sigma^2,
}
where $m_\sigma(\phi_\star) =M_0$ is the minimum mass, $\mu = g\phi_\star$, and $m_\sigma(\phi_\star)m_\sigma^{\prime\prime}(\phi_\star) = g^2$ is a coupling parameter.
Using these and the slow-roll expansion $\phi-\phi_\star \approx \dot{\phi}_0(t-t_\star)$, we can rewrite Eq.~\eqref{eq:eff_mass} as
\es{eq:eff_mass_2}{
m_\sigma(\phi) \approx M_0 + \frac{g^2}{2M_0}\dot{\phi}_0^2(t-t_\star)^2,
}
where the slow-rolling velocity is given by $\dot{\phi}_0 \approx (60 H)^2$, from the normalization of the CMB anisotropy power spectra. For $g\sim {\cal O}(10)$ and $M_0 \sim {\cal O}(100)H_I$, as will be the focus of our search, the time-varying second term dominates over the first in Eq.~\eqref{eq:eff_mass_2}.
This shows that for a narrow window in time around $t_\star$, $\sigma$ becomes significantly lighter than usual, and $\sigma$ particle production can occur.
The number density of produced particles can be computed using standard techniques~\cite{Kofman:1997yn}.

\begin{figure}
    \centering
    \includegraphics[width=0.48\textwidth]{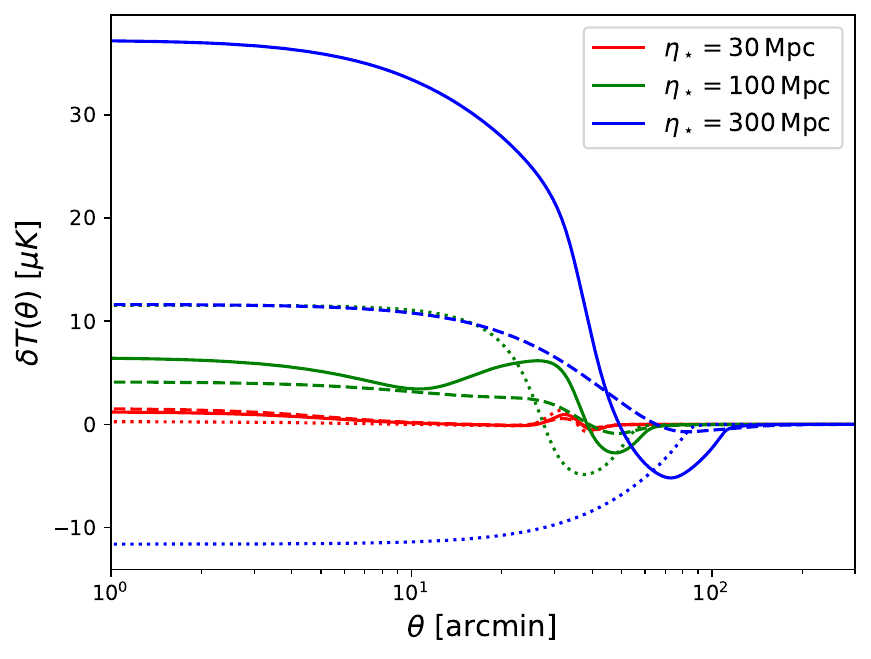}
    \caption{Characteristic examples of the particle-production hotspot profiles. We show profiles as a function of opening angle $\theta\equiv \mathrm{cos}^{-1}(\hn\cdot\hn_{\rm HS})$ for three values of the horizon size at production $\eta_\star$ (indicated by color) and three values of the hotspot distance $\eta_{\rm HS}$ ($\mathrm{max}(\eta_{\rm rec}-\eta_\star,0)$ in dotted lines, $\eta_{\rm rec}$ in full lines, and $\eta_{\rm rec}+\eta_\star$ in dashed lines). Whilst the individual shapes differ considerably, it is clear that the profile amplitude increases roughly linearly with $\eta_\star$. Each profile is computed using Eq.~\eqref{eq: template-def}, without an instrumental beam.}
    \label{fig: profiles}
\end{figure}

The time-varying large mass gives rise to a gravitational potential around the location of each produced particle, sourcing a non-zero one-point function of the curvature perturbation~\cite{Maldacena:2015bha}.
One way to understand this is to note that Eq.~\eqref{eq:lag} determines an interaction between the inflaton and $\sigma$.
The produced $\sigma$ particles can exert a force on the inflaton, slowing down the inflaton evolution around their locations.
This means inflation ends later in those localized regions compared to usual, giving rise to overdensities (noting that regions where inflation ends earlier experience more post-inflationary dilution of energy density, and thus end up as underdensities). When CMB decoupling happens these overdensities could give rise to either hot or cold spots,\footnote{For concision, we will frequently refer to both possibilities as ``hotspots.''} depending on the combination of Sachs-Wolfe, Doppler, or integrated Sachs-Wolfe effects.
Detailed computation of the temperature profile of these spots, as well as their distribution in the sky can be found in~\cite{Kim:2021ida, Kim:2023wuk}.
In particular, Refs.~\citep{Kim:2021ida,Kim:2023wuk} showed that a shell of thickness $\Delta\eta$ around the surface-of-last-scattering would be expected to contain
\beq\label{eq: N-spots}
    N_{\rm spots} &\approx& 4 \times 10^8\times g^{3/2}\left(\frac{\Delta\eta}{100\,\mathrm{Mpc}}\right)\left(\frac{100\,\mathrm{Mpc}}{\eta_\star}\right)^3\\\nonumber
    &&\,\times\,e^{-\pi(M_0^2-2H_I^2)/(g|\dot\phi_0|)}
\eeq
hotspots that depend on the size of the comoving horizon $\eta_\star$ at the time of particle production, involving the usual exponential dependence on the (minimum) mass $M_0$. Furthermore, a single event at distance $\chi_{\rm HS}\equiv \eta_0-\eta_{\rm HS}$, where $\eta_0$ is the present day comoving horizon and $\eta_{\rm HS}$ is the comoving location of the spot, and angular position $\hn_{\rm HS}$ is expected to yield the temperature profile:\footnote{We drop contributions with $\ell<2$ since these are degenerate with the mean CMB temperature and dipole, and thus difficult to observe in practice.}
\beq\label{eq: template-def}
    \delta T(\hn;\hn_{\rm HS},\eta_\star,\eta_{\rm HS}) &=& \frac{1}{2\pi^2}\sum_{\ell=2}^\infty (2\ell+1)\mathcal{L}_\ell(\hn\cdot\hn_{\rm HS})\\\nonumber
    &&\,\times\,\int_{0}^{\infty}\frac{dk}{k} j_\ell(k\chi_{\rm HS})\mathcal{T}_\ell(k)f(k\eta_\star).
\eeq
Here $f(x) = gH_I^2/\dot{\phi}_0\times \left[\mathrm{Si}(x)-\sin(x)\right]$
(for Sine integral $\mathrm{Si}$), $\mathcal{T}_\ell(k)$ is the CMB transfer function (including Sachs-Wolfe, integrated Sachs-Wolfe, and Doppler contributions), $j_\ell$ is a spherical Bessel function, and $\mathcal{L}_\ell$ is a Legendre polynomial. Importantly, this is linearly proportional to the coupling amplitude $g$: this fact forms the backbone of our estimation pipeline. In Fig.~\ref{fig: profiles}, we show a set of exemplar temperature profiles, highlighting their phenomenological variability.

\section{Methodology}
\noindent As discussed in \citep{Kim:2021ida,Kim:2023wuk}, the inflationary model of the previous section generates \textit{pairwise} hotspots, separated by some distance $\lesssim \eta_\star$. Performing an optimal analysis for a pair of sources is difficult, since the necessary template is anisotropic with several degrees of freedom. To avoid this difficulty, we here perform a search using only a single (isotropic) hotspot template, which \citep{Kim:2023wuk} find to be close-to-optimal in practice. In particular, we perform a matched-filter analysis to constrain the coupling parameter $g$, which scales the primordial signal given in Eq.~\eqref{eq: template-def}. To implement this, we follow a similar procedure to the \textit{Planck} tSZ cluster searches \citep{Planck:2013shx,Planck:2015koh}, using a modified version of the \textsc{szifi} code described in \citep{Zubeldia:2022mvz,Zubeldia:2022gva}. 

In full, our pipeline involves the following steps, applied to the six \textit{Planck} HFI intensity maps (from $100$ to $857$ GHz):
\begin{enumerate}
    \item Divide the sky into 768 square cut-outs (of size $14.8\,\mathrm{deg}\times 14.8\,\mathrm{deg}$) tiling the entire sky, each containing $1024^2$ pixels (following \citep{Zubeldia:2022mvz}).
    \item Mask the Galactic plane using the \textit{Planck} GAL090 mask, as in \citep{Planck:2015koh}, and inpaint any point sources (see below) using a diffusive algorithm. The product of the inpainted frequency maps, the Galactic mask, and the projection mask defines the data, $d_\nu(\vtheta)$. 
    \item For each cut-out, compute the filtered map $\hat{g}(\vtheta)$ via a 2D convolution (here for $l\in[30,3000]$)~\cite{Haehnelt:1995dg,Herranz:2002kg,Melin:2006qq}:
    \beq\label{eq: matched-filter}
        \hat{g}(\vtheta) &=& \sigma_g^2\int\frac{d^2\vec l}{(2\pi)^2} t_{\nu}^*(\vec l;\vtheta)\mathsf{C}^{-1,\nu\nu'}(l)d_{\nu'}(\vec l)\\\nonumber
        \sigma_g &=& \left(\int\frac{d^2\vec l}{(2\pi)^2} t_{\nu}^*(\vec l;\vec 0)\mathsf{C}^{-1,\nu\nu'}(l)t_{\nu'}(\vec l;\vec 0)\right)^{-1/2},
    \eeq
    where $t_\nu(\vl;\vtheta)$ is the Fourier-space hotspot template centered at $\vtheta$, and $\mathsf{C}_{\nu\nu'}(l) = \av{d_\nu(\vec l)d^*_{\nu'}(\vec l)}$ is the diagonal-in-$l$ covariance (including primary, secondary, and noise contributions), which is estimated from the data directly. Here, $\hat{g}$ is an unbiased estimator of the coupling parameter $g$, whose variance (which depends on the cut-out of interest) is given by $\sigma^2_g$ \citep{Zubeldia:2022mvz}. Note that Eq.~\eqref{eq: matched-filter} synthesizes the data from all six HFI intensity maps into a single $\hat{g}$ map, with the SED of the hotspot signal identical to that of the blackbody primary CMB, which is unity at all frequencies in the thermodynamic CMB temperature units employed here.
    \item Identify a hotspot as any region of the map with $\hat{g}(\vtheta)\geq 5\sigma_g$ using a density-based spatial clustering algorithm \citep{dbscan}.
    \item Repeat steps 3 and 4 for all hyperparameters ($\eta_\star$ and $\eta_{\rm HS}$) of interest. For any detections, assign them the hyperparameter set with the largest detection significance (\textit{i.e.}, $\hat{g}/\sigma_g$).
    \item Apply cuts, as described below, to create a final catalog of hotspots across all tiles.
\end{enumerate}
In this work, we consider $10$ choices of $\eta_\star$ logarithmically spaced in $[10,1000]\,\,\mathrm{Mpc}$, as well as $10$ linearly spaced $\eta_{\rm HS}$ values, each satisfying $\chi_{\rm HS} \in [\chi_{\rm rec}-\eta_\star,\chi_{\rm rec}+\eta_\star]$ and $0\leq \chi_{\rm HS}\leq \eta_0$, for comoving distance $\chi(\eta) \equiv \eta_0-\eta$. \resub{The conditions on $\eta_{\rm HS}$} ensure that the particle production events lie in the observable Universe, and directly affect the last-scattering surface. \resub{Furthermore, the lower limit on $\eta_\star$ is set by the \textit{Planck} beam: hotspots with $\eta_\star\lesssim 10\,\mathrm{Mpc}$ cannot be distinguished from point sources, except through their frequency dependence.}

To generate the hotspot templates, we first implement the profile of Eq.~\eqref{eq: template-def} (with $g=1$), using $\ell_{\rm max}=3500$ and \textsc{camb}~\cite{Lewis:1999bs}\footnote{\url{https://camb.info/}} transfer functions defined across $1000$ logarithmically-spaced $k$-points in $[10^{-6},1]\,\,\Mpch$. This is computed on the $1024^2$ flat-space pixel grid (using the flat-space distance $\theta = \mathrm{cos}^{-1}(\hn\cdot\hn_{\rm HS})$), and numerically convolved with the isotropic \textit{Planck} beam for each frequency of interest~\cite{2016A&A...594A...7P}. For efficiency, and to avoid the edges of the cut-out tiles, we limit to $\theta<\theta_{\rm max} = 0.1$~radian ($\approx 5\,\,\mathrm{deg}$) or, if larger, the scale at which the profile is less than $1\%$ of the peak; this practically restricts us to $\eta_\star \lesssim 1000$~Mpc. 

We apply a number of cuts to limit contamination of primordial hotspots from late-time contributions. For \textit{Planck} analyses, one could directly utilize public catalogs for this purpose \citep[{\it e.g.},][]{Planck:2015bin,Planck:2015koh}; to ensure consistency with the simulation-based tests below, we here measure them directly from the dataset (though we additionally compare any \textit{Planck} detection candidates to the public catalogs in the below). We first find all point sources with $\mathrm{SNR}\geq 10$ in any of the six frequency channels, using a modified version of the above algorithm (analyzing each dataset in turn with a beam-convolved point source template). The union of these catalogs defines a full-sky point source map (here computed at $N_{\rm side}=4096$), which we use as a mask, excising regions within $3\sigma_{{\rm beam},\nu}$ via a recursive search, given the beam-size $\sigma_{{\rm beam},\nu}$ in channel $\nu$ (which increases with decreasing frequency, from roughly $2'$ at 857 GHz to $4'$ at 100 GHz).\footnote{Recall that $\sigma_{\rm beam} = {\rm FWHM}_{\rm beam}/\sqrt{8 \ln 2}$, where FWHM is the full-width at half-maximum.} \resub{Whilst most \textit{Planck} analyses exclude all point sources with $\mathrm{SNR}\geq 5$, we use a less conservative choice in this work matching the \textit{Planck} tSZ cluster analysis \citep{Planck:2015koh}. This ensures that we do not `over-mask' and miss any hotspot candidates; to check for residual contamination, we compare the hotspot catalog to the set of $\mathrm{SNR}\geq 5$ point sources in the post-processing step.}

Secondly, we perform a matched-filter analysis for tSZ clusters following \citep{Planck:2015koh}, using the pipeline above, as implemented in \textsc{szifi} \citep{Zubeldia:2021mzx,Zubeldia:2022gva,Zubeldia:2022mvz}. This uses a tSZ SED in the MMF of Eq.~\eqref{eq: matched-filter}, and $15$ values of the cluster size, $\theta_{500}\in[5,50]'$ (for the pressure profile of~\cite{Arnaud2010}). Following \citep{Zubeldia:2022mvz}, we compute the SNR iteratively, masking out previously detected clusters when estimating the data covariance $\mathsf{C}_{\nu\nu'}(l)$. Here, templates are applied for $l\in[100,2500]$ (matching previous works) and we apply an SNR threshold of $4.5$ (as in \citep{Planck:2013shx,Planck:2015koh}), additionally dropping any clusters within $5\sigma_{{\rm beam},\nu}$ of a point source. This catalog is used to post-process the hotspot catalog \resub{by rejecting any candidates within $10'$ of a tSZ cluster. This approach is, by construction, conservative: we consider only hotspot candidates that are not flagged by the point source or tSZ searches, rather than attempting to distinguish between the various types of source.}

Finally, we apply a conservative Galactic mask to the output catalog, removing any potential sources within the brightest $\approx 40\%$ of the sky in the highest frequency channel. 
This helps to avoid contamination from dusty clumps in the Milky Way and Magellanic Clouds, as described in the \textit{Planck} Compton-$y$ map paper \citep{Planck:2015vgm}. In concert with the above procedures, this leaves a total of $57.2\%$ of the sky for analysis (straightforwardly computed by injecting points and calculating the fraction not removed by masking operations). The end result of these procedures is a catalog of hotspot candidates, which can then be used for analysis (or for validation, if one injects a set of fake hotspots). An additional check (which we consider in the \textit{Planck} section below) is to repeat the analysis using a standard (non-multi-frequency) matched filter applied to a component-separated CMB temperature map, rather than applying an MMF to the frequency maps; assuming an effective component separation method, this \resub{further reduces contamination from tSZ clusters and point sources, but should yield} comparable SNR, since the hotspots have a blackbody SED.

\section{Validation Tests}
\noindent We now proceed to test the above methodology using synthetic data. For this purpose, we use a single \textit{Planck} FFP10/\textsc{npipe} simulation \citep{Planck:2020olo},\footnote{Available on NERSC: \\\href{https://portal.nersc.gov/project/cmb/planck2020/}{https://portal.nersc.gov/project/cmb/planck2020/}.}, taking the six highest frequencies, as above. We consider two recovery tests: (1) injecting single hotspots at known $\eta_\star$ and fixed $\eta_{\rm HS}$; (2) injecting pairwise hotspots at known $\eta_\star$ with a distribution of separations and $\eta_{\rm HS}$ values. The first validates our pipeline under ideal conditions, whilst the second tests whether it remains efficacious in the desired primordial use-case.

\subsection{Single Hotspots}

\begin{figure*}
    \centering
    \includegraphics[width=0.32\textwidth]{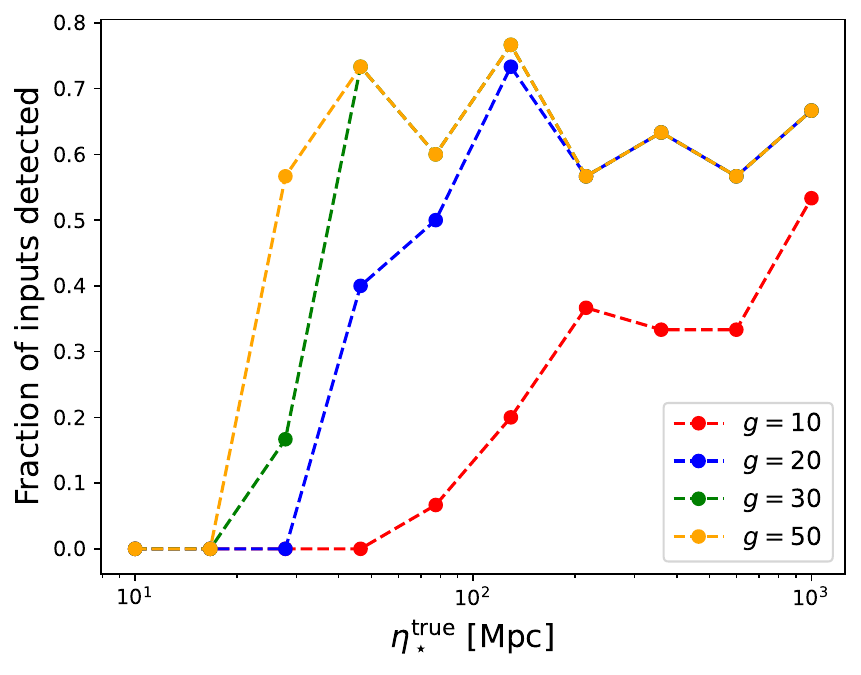}
    \includegraphics[width=0.32\textwidth]{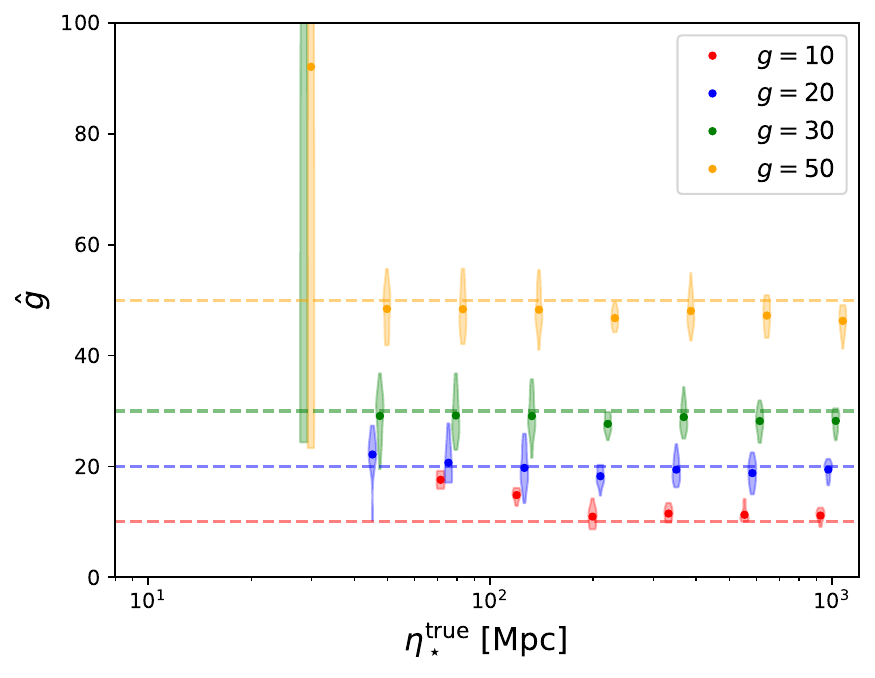}
    \includegraphics[width=0.32\textwidth]{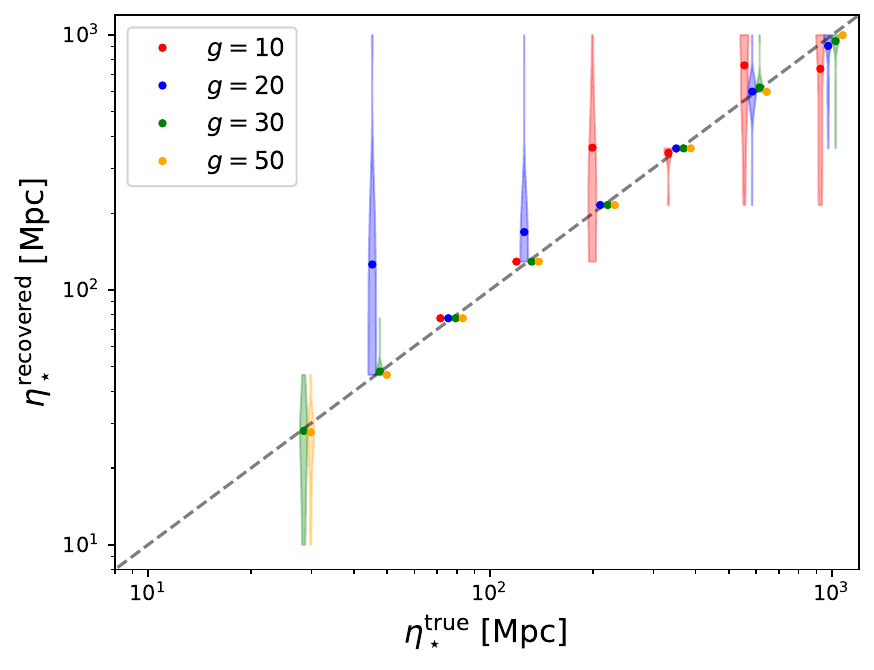}
    \caption{Validation tests of the hotspot-finding pipeline, as applied to a single FFP10/\textsc{npipe} simulation contaminated with $300$ isotropic hotspots across ten values of the size parameter $\eta_\star$, with distance $\eta_{\rm HS}$ fixed to $\eta_{\rm rec}$. \textbf{Left panel}: fraction of injected sources recovered at $\gtrsim 5\sigma$ as a function of the coupling parameter $g$. This asymptotes to $\approx 60\%$ due to masking, reflecting the fraction of sources obscured by late-time effects and Galactic contamination. Relative to the masked input catalog, the completeness is $100\%$ for sufficiently large $g,\eta_\star$. Larger hotspots are easier to detect: we are limited on small scales by the \textit{Planck} beam, and on large scales by the flat-sky assumption and the Galactic mask. \textbf{Middle panel}: estimated values of $g$ from the injected sources. Though the uncertainty is large for small $\eta_\star$ (where hotspots are difficult to detect, with no sources for $\eta_\star<25\,\,\mathrm{Mpc}$), we find generally unbiased results at larger $\eta_\star$. \textbf{Right panel}: estimated values of $\eta_\star$ from the injected sources. As before, these are uncertain for small sources, but converge for large $g, \eta_\star$. We add small displacements along the abscissa for clarity.}
    \label{fig: single-test}
\end{figure*}

\noindent We begin by creating a map of $300$ hotspots at unit $g$, isotropically distributed on the full sky. Via Eq.~\eqref{eq: template-def}, hotspots are defined on an $N_{\rm side}=2048$ \textsc{HEALPix} map, convolved with the \textit{Planck} beam as before. For simplicity, we fix $\eta_{\rm HS}=\eta_{\rm rec}$ (\textit{i.e.}, place hotspots on the last-scattering surface \citep[cf.,][]{Kim:2021ida}) but use $10$ values of $\eta_\star$ (each with $30$ hotspots). To avoid confusion in the inference, we ensure that all hotspots are separated by at least $1\,\,\mathrm{deg}$, such that they can be treated independently. This map is then combined linearly with the FFP10/\textsc{npipe} temperature map at each frequency for $g\in[10,20,30,50]$ (following initial testing). \resub{Given an analysis mask (built from the uninjected map to avoid the hotspots saturating the point source algorithm and thus being subtracted)}, we search for hotspots, using the matched-filter pipeline: using $10$ $\eta_\star$ choices, this requires $\approx 20$ CPU-hours for each value of $g$, and can be trivially parallelized. 

In Fig.~\ref{fig: single-test}, we show the main results of the simulated single-hotspot analysis: the properties of injected hotspots recovered by our pipeline as a function of their (true) size, $\eta_\star^{\rm true}$.\footnote{We declare a hotspot to be ``recovered'' if there is a $\geq 5\sigma$ candidate whose center lies within $10'$ of the true center. If there are multiple (which commonly occurs around noisy hotspots), we take parameters from the candidate with largest SNR.
} 
From the first plot, we observe that our pipeline achieves good completeness for sufficiently large $g$ and $\eta_\star$; relative to the unmasked catalog of inputs, we find a recovery rate of $(64\pm7)\%$ for $\eta_\star>25\,\,\mathrm{Mpc}$ and $g>20$, with the remaining hotspots hidden by the analysis mask \resub{(which removes the Galactic plane and known secondary sources, such as tSZ clusters)}. If one instead defines the completeness with respect to the \textit{masked} catalog, this increases to $100\%$, such that we recover all the inputs. As expected, hotspots with larger $g$ are easier to detect, with the completeness saturating at $g\approx 20$ for $\eta_\star>100\,\mathrm{Mpc}$. As $\eta_\star$ decreases, the completeness reduces significantly; small sources yield small temperature perturbations \citep[cf.,][]{Kim:2021ida} and are \resub{swamped by the \textit{Planck} detector noise; moreover, they would be indistinguishable from point sources, except through their frequency dependence}. Hotspots with $\eta_\star>1000\,\,\mathrm{Mpc}$ (and angular sizes $\gtrsim 5\,\,\mathrm{deg}$) may also be detectable, though these are difficult to analyze since one cannot assume the flat-sky limit, may suffer from significant Galactic-plane contamination, and primary CMB cosmic variance is large on these scales.

Fig.~\ref{fig: single-test} also shows the recovered parameters of the injected hotspots: $\hat{g}$ and $\eta_\star^{\rm recovered}$. For sufficiently large hotspots, we report generally unbiased constraints on both parameters, though substantial uncertainty in $\eta_\star^{\rm recovered}$ for low-completeness samples. For large coupling amplitudes, the estimated amplitudes $\hat{g}$ are slight underestimates; this occurs since the hotspots contribute non-trivially to the measured covariance of the data, and can be mitigated using iterative analyses \citep{Zubeldia:2022mvz}. 

The conclusion of this exercise is that our approach can robustly extract and constrain hotspots (using a $5\sigma$ threshold) with $g\gtrsim 10$ for $\eta_\star\gtrsim 100\,\,\mathrm{Mpc}$. With a higher-precision experiment, one expects significantly higher sensitivity to small sources, but little change for large $\eta_\star$, given that \textit{Planck} is already cosmic-variance-limited at $\ell \lesssim 1600$ in temperature.\footnote{Further constraints could be wrought from CMB polarization, for which \emph{Planck} is not cosmic-variance-limited.} Notably, the inferred bounds on $g$ are considerably weaker than those claimed in \citep{Kim:2023wuk} (which applied matched filters to Gaussian random field data); following some analysis, we have concluded that the former constraints were overconfident since their hotspot templates included cuts in real space (leading to excess power on small scales in Fourier space) and did not include an experimental beam or noise.

\subsection{Pairwise Hotspots}

\begin{figure*}
    \centering
    \includegraphics[width=0.47\textwidth]{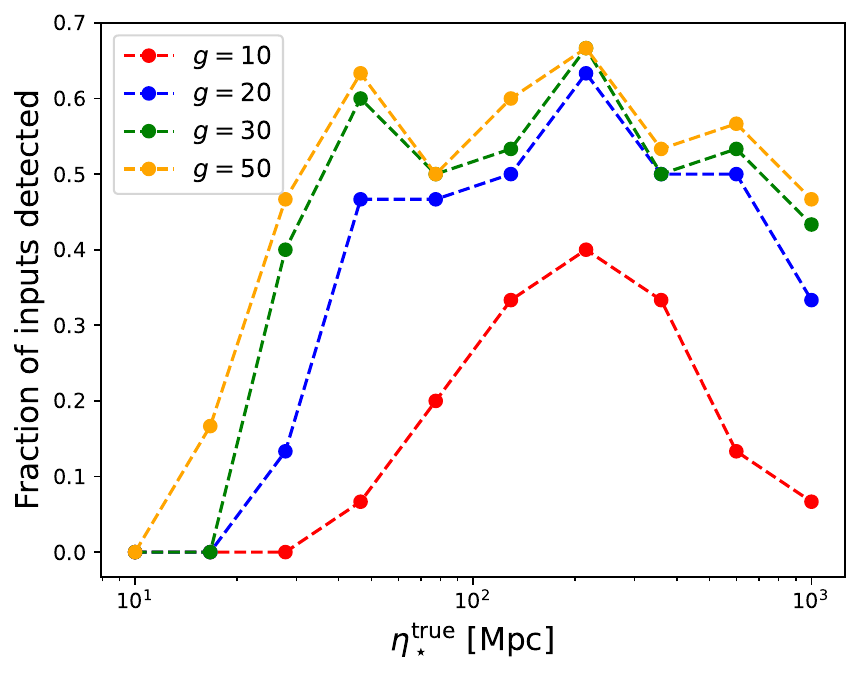}
    \includegraphics[width=0.47\textwidth]{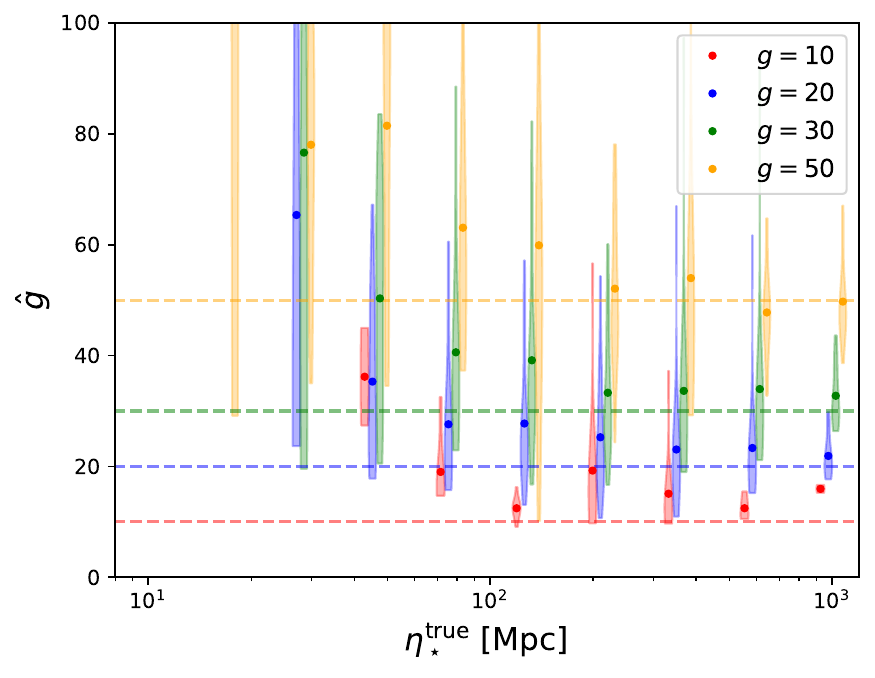}\\
    \includegraphics[width=0.47\textwidth]{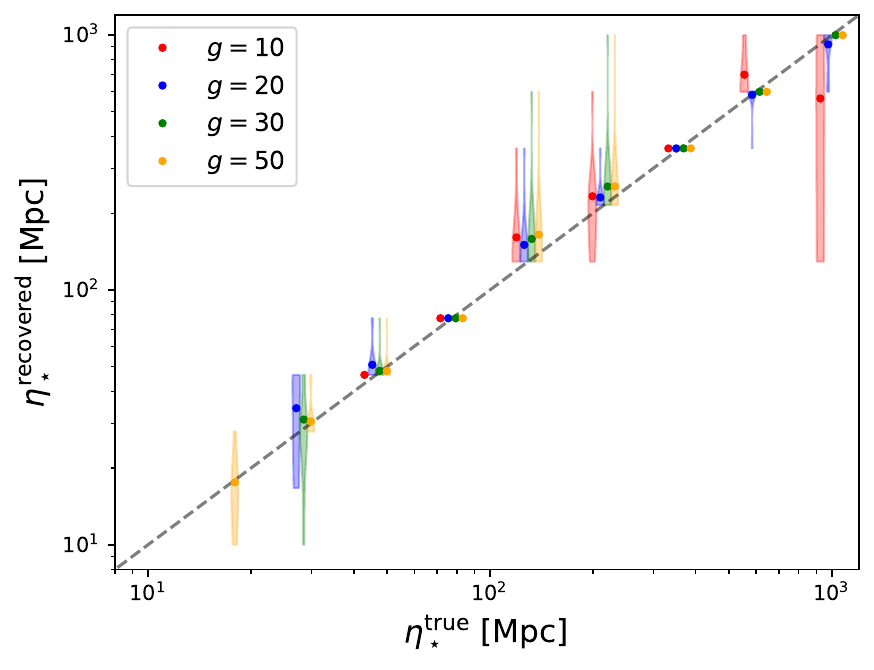}
    \includegraphics[width=0.47\textwidth]{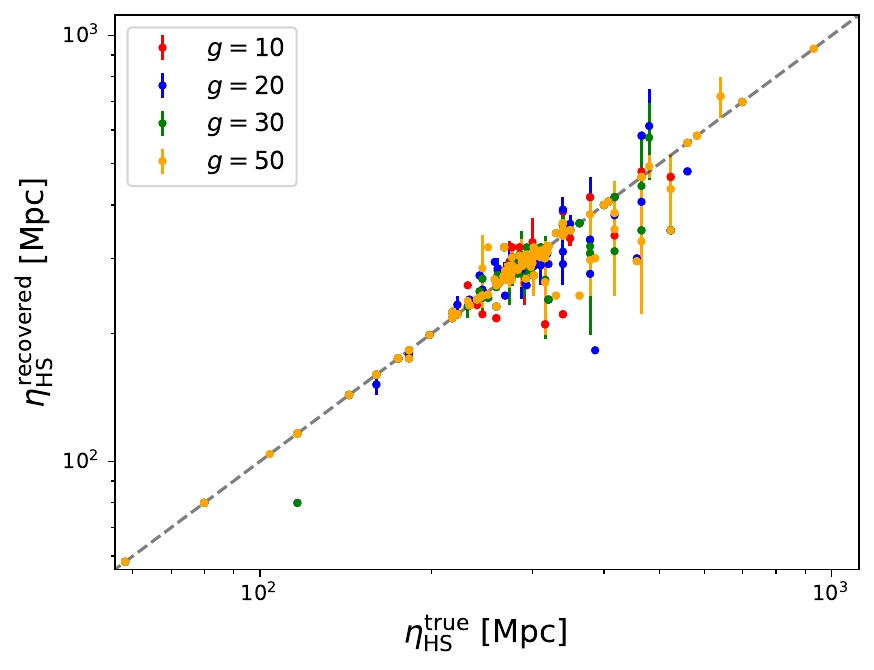}
    \caption{As Fig.\,\ref{fig: single-test}, but validating the pipeline on a simulation contaminated by $300$ \textit{pairs} of hotspots. We use ten values of $\eta_\star$ as before, but now vary also the hotspot distances $\eta_{\rm HS,1/2}$ with parameters drawn from uniform distributions as described in the text. Due to the overlapping signals, the pairwise hotspots are easier to detect at low $\eta_\star$ (evidenced by higher completenesses, see the upper left panel, noting that we remain limited by the mask on large scales), though this overlap also leads to positively biased values of $\hat{g}$ (upper right panel). As for the single hotspots, the size parameters are recovered well (lower left panel), though the errors increase slightly due to hotspot confusion. The lower right panel shows the recovered values of $\eta_{\rm HS}$, combining results from both hotspots. These lie close to the input values, giving further validation of the pipeline.}
    \label{fig: pair-test}
\end{figure*}

\noindent Next, we generate a map of pairwise hotspots, following a similar procedure to \citep{Kim:2023wuk}, though generalized outside the flat-sky limit. We first draw a single hotspot position $(\hn_{\rm HS,1},\chi_{\rm HS,1})$ given the aforementioned bounds, using a uniform distribution in Cartesian space. A second point is then drawn from a ball of radius $\eta_\star$ about the first (uniformly in Cartesian coordinates), and the pair is rejected if $\chi_{\rm HS,2}$ is not within $\eta_\star$ of the recombination surface (or outside the observable Universe). This procedure is repeated to generate $300$ pairs of points with known $\eta_{\rm HS,1,2}$ (which is discretized to match the values used in the analysis), including $10$ values of $\eta_\star$. These are used to construct a full-sky template map from Eq.~\eqref{eq: template-def}, which is co-added to the FFP10/\textsc{npipe} simulation, as before. The resulting map is then analyzed via the above pipeline. Importantly, the synthetic dataset contains hotspot pairs whilst the matched filter assumes single hotspots: this approach tests whether our method is applicable in the theory-motivated scenario of a distribution of pairs.

The results of this test are summarized in Fig.~\ref{fig: pair-test}. For small hotspots (low $\eta_\star$) we find somewhat enhanced completeness compared to the single-spot case shown in Fig.~\ref{fig: single-test}, though the method performs slightly worse at high $\eta_\star$, now with an asymptotic value of $50-60\%$ (or $90-100\%$ when accounting for masking). This may be justified by noting that small overlapping hotspots create larger temperature perturbations, which can be more easily recovered with our pipeline. For large $\eta_\star$, we allow for significant spread in the values of $\eta_{\rm HS}$ (within $[\eta_{\rm rec}\pm\eta_\star]$, also assuming $0\leq \eta_{\rm HS}\leq \eta_0$), which can lead to destructive inference (cf.~Fig.~\ref{fig: pair-test}). Furthermore, it is more difficult to find the center of an extended overlapping region, which may result in some large-$\eta_\star$ sources being missed (recalling that we claim a detection if the true and injected hotspots lie within $10'$). The above notwithstanding, it is clear from the figure that our approach can find hotspots at $g\gtrsim 10$ with high efficacy.

The remaining panels of Fig.~\ref{fig: pair-test} show the recovered hotspot hyperparameters. In contrast to Fig.~\ref{fig: single-test}, the inferred coupling strength $\hat{g}$ incurs some positive bias; this is caused by the overlap of hotspots, and is a consequence of using a different template for the dataset generation and analysis (by necessity). That said, the recovered values of $\eta_\star$ are in good agreement with the inputs (albeit with increased scatter), and we additionally find excellent recovery of the hotspot distances, $\eta_{\rm HS}$, across all values injected ($10$ for each $\eta_\star$ choice). The latter observation affords us confidence that our pipeline can be applied to situations when both $\eta_\star$ and $\eta_{\rm HS}$ are varied (which was not considered in the previous subsection).

We comment that in the context of the model in Eq.~\eqref{eq:lag}, values of $g\gg \sqrt{4\pi}$ are non-perturbatively large.
However, in scenarios with \resub{an extradimensional model of inflation, an ``effective'' value of $g\sim 10$ can be realized without violating such perturbativity bounds~\cite{Kumar:2025gon}} whilst maintaining the same phenomenological hotspot properties.
\resub{Scenarios involving tachyonic Higgs production~\cite{Shakya:2023zvs} and quantum diffusion effects~\cite{Ezquiaga:2022qpw} may also lead to similar signatures.}
Therefore, to be model-agnostic we consider $g\sim 10-100$ here.

\section{Application to Planck}
\noindent Finally, we apply our pipeline to the observed \textit{Planck} dataset. For this purpose, we use the \textit{Planck} PR4 HFI temperature maps \citep{Planck:2020olo,Tristram:2020wbi,Carron:2022eyg,Rosenberg:2022sdy}, in combination with the point source and Galactic masks described above. We search for hotspots using $100$ templates across $10$ logarithmically-spaced values of $\eta_\star$ and $10$ linearly-spaced values of $\eta_{\rm HS}$, obtaining a catalog of detection candidates as before. This requires $\sim 250$ CPU-hours in total. To account for any candidates appearing multiple times in the catalog, we merge any pair separated by less than the mean hotspot radius ($\approx\sqrt{4\pi}(\eta_\star^1+\eta_\star^2)/2\eta_0)$, \citep{Kim:2021ida}). To validate any potential detections, we additionally perform an analysis using the \textsc{sevem} and \textsc{commander} component-separated maps \citep{Planck:2020olo}. This proceeds as above, but effectively involves only a single-frequency matched filter rather than MMF, and additionally uses the \textit{Planck} common component-separation mask, \resub{which excludes any point sources with $\mathrm{SNR} \geq 5$} \citep{Planck:2018yye}. 

\begin{figure*}
    \centering
    \includegraphics[width=0.8\textwidth]{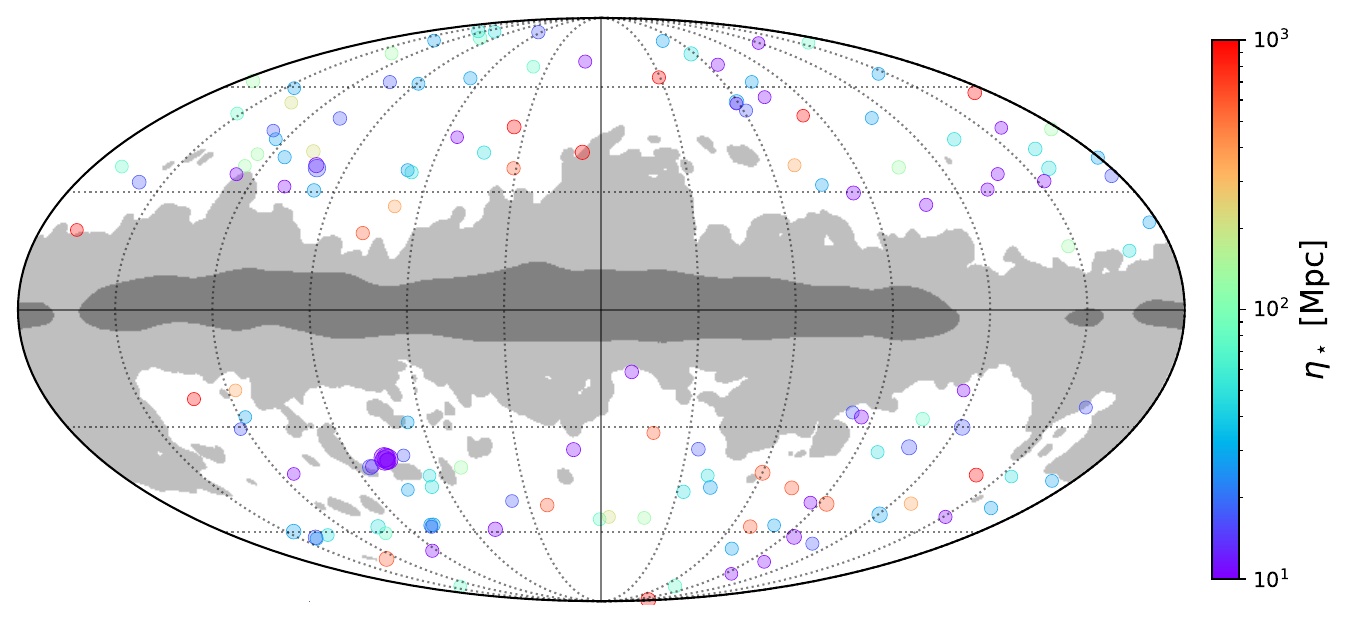}
    \caption{Visualuzation of the hotspot detection candidates from the full \textit{Planck} PR4 high-frequency data. The dark and light shaded regions show the region masked out by the analysis mask and post-processing mask respectively. Each circle represents a single detection candidate with $\mathrm{SNR}\geq5$, whose size is proportional to the SNR and whose color indicates the inferred value of $\eta_\star$. The dark purple circle in the south-west corresponds to multiple similarly-located candidates; as discussed in the text, this is likely due to an unsubtracted compact source. A histogram of the detection significances is shown in Fig.~\ref{fig: planck-det}, with the parameters of the highest-SNR detections given in Tab.~\ref{tab: candidates}.}
    \label{fig: planck-map}
\end{figure*}

\begin{figure}
    \centering
    \includegraphics[width=0.49\textwidth]{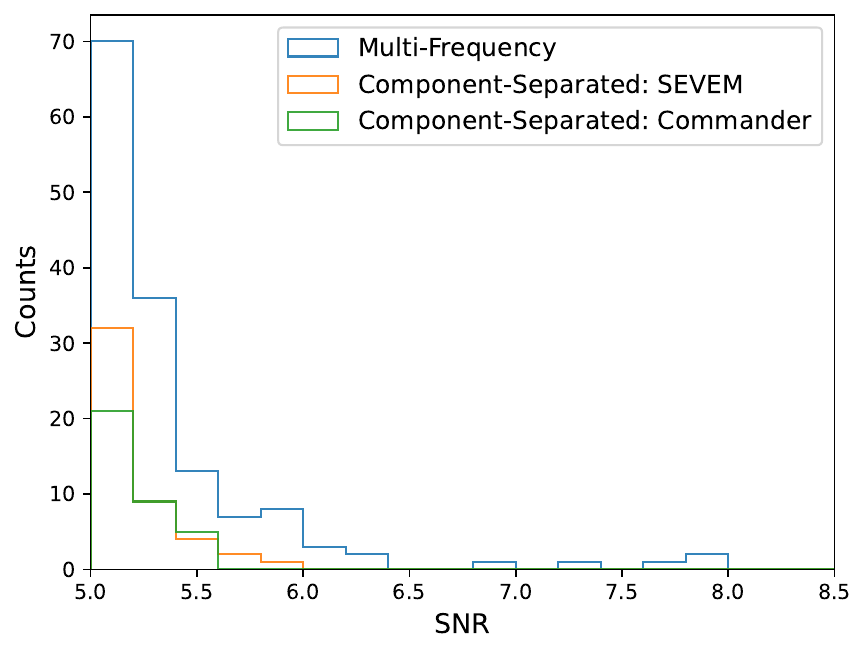}
    \caption{PDF of \textit{Planck} PR4 detections at $\mathrm{SNR}\geq 5$. We compare results from the full six HFI channels (blue) and those from component-separated maps (orange, green) using two pipelines. These detections appear compatible with noise, given that the $\mathrm{SNR}\geq 6$ detections are not found in the component-separated pipeline.}
    \label{fig: planck-det}
\end{figure}

In Fig.~\ref{fig: planck-map}, we show the distribution of $\mathrm{SNR}\geq5$ hotspot candidates across the sky. Outside the mask, these appear roughly isotropically distributed, with an array of different $\eta_\star$ values. This is good: we do not find that all detection candidates cluster in regions of higher Galactic contamination. A more useful diagnostic is the distribution of SNR values, which is shown in Fig.~\ref{fig: planck-det}. In total, we find $144$ candidates, with SNR up to $8$; it is important to note that our analysis has four free parameters ($\eta_\star$, $\eta_{\rm HS}$, and the hotspot center coordinates), thus moderately high SNR values are expected due to chance fluctuations. Furthermore, analysis of the \textsc{sevem} (\textsc{commander}) component-separated map finds only 48 (35) candidates, with a maximum detection significance of $6\sigma$. This indicates that the $\mathrm{SNR}\geq 6$ candidates (which all have $\mathrm{SNR}<5$ in component-separated maps) may be spurious instead of primordial.

\begin{table}[]
    \centering
    \begin{tabular}{c|ccccc}
    SNR & Longitude [$^\circ$] & Latitude [$^\circ$] & $g$ & $\eta_\star$ [$\mathrm{Mpc}$] & $\eta_{\rm HS}$ [$\mathrm{Mpc}$]\\\hline
    8.0&76.7&-38.6&631&10.0&272.2\\
    7.9&77.8&-38.1&692&10.0&277.7\\
    7.7&78.0&-38.9&651&10.0&275.9\\
    7.3&77.1&-38.1&580&10.0&272.2\\
    6.9&100.3&36.3&98&16.7&266.8\\
    6.4&77.4&-39.1&698&10.0&288.6\\
    6.3&76.8&-38.9&494&10.0&272.2\\
    6.2&238.3&-30.0&134&16.7&281.9\\
    6.1&239.3&-54.6&34&27.8&257.7\\
    6.0&84.8&-40.7&114&16.7&266.8\\
    \end{tabular}
    \caption{\textit{Planck} hotspot candidates ordered by SNR. For each row, we give the inferred parameters of a hotspot candidate found in the \textit{Planck} PR4 HFI dataset. Longitude and latitude are specified in Galactic coordinates. Note that (a) none of these candidates are found in the component-separated maps, (b) all the candidates have $\eta_\star<30\,\mathrm{Mpc}$ (which show the lowest completeness in the injection tests), and (c) six of these candidates are closely separated on the sky. Visual inspection of each candidate can be found in Fig.~\ref{fig: inspection}.}
    \label{tab: candidates}
\end{table}

Tab.~\ref{tab: candidates} lists the parameters of the six-frequency \textit{Planck} $\mathrm{SNR}\geq 6$ hotspot candidates, with visualizations of the \textit{Planck} maps around each point shown in Fig.~\ref{fig: inspection}. Each corresponds to a hotspot with temperature perturbation around $30\,\,\mu$K. Notably, all potential hotspots are small, with $\eta_\star<30\,\,\mathrm{Mpc}$ in all cases; as shown previously, physical hotspots of this size are the hardest to detect due to the \resub{increased noise on small-scales, and the broad \textit{Planck} beam, which leads to small hotspots having the same spatial signatures as point sources -- this is an important limitation of the method, which can be improved with the use of higher-resolution CMB data.} Furthermore, six of the ten sources are located in a ring of radius $\sim 1^\circ$ and have similar inferred parameters, which suggests a ringing signature around a correlated source. Indeed, cross-matching with the public \textit{Planck} catalogs (including point sources, Galactic cold clumps, tSZ clusters, and high-redshift sources), we find a non-thermal source (\textit{i.e.}, synchrotron) \citep{Planck:2018kuk} at the center of the ring, which is clearly seen in Fig.~\ref{fig: inspection}. We additionally find a non-thermal source within $10'$ of the $\mathrm{SNR}=6.0$ candidate and a $5\sigma$ point source within $13'$ of the $\mathrm{SNR}=6.9$ candidate. All remaining sources are excluded by the \textit{Planck} GAL020 mask (though this represents a harsh foreground cut), with the $\mathrm{SNR}=6.2$ candidate masked in the \textit{Planck} inpainting map. All-in-all, we find no significant evidence for a hotspot with $\mathrm{SNR}\geq 6$ in the \textit{Planck} data.

\begin{figure*}
    \centering
    \includegraphics[width=0.49\textwidth]{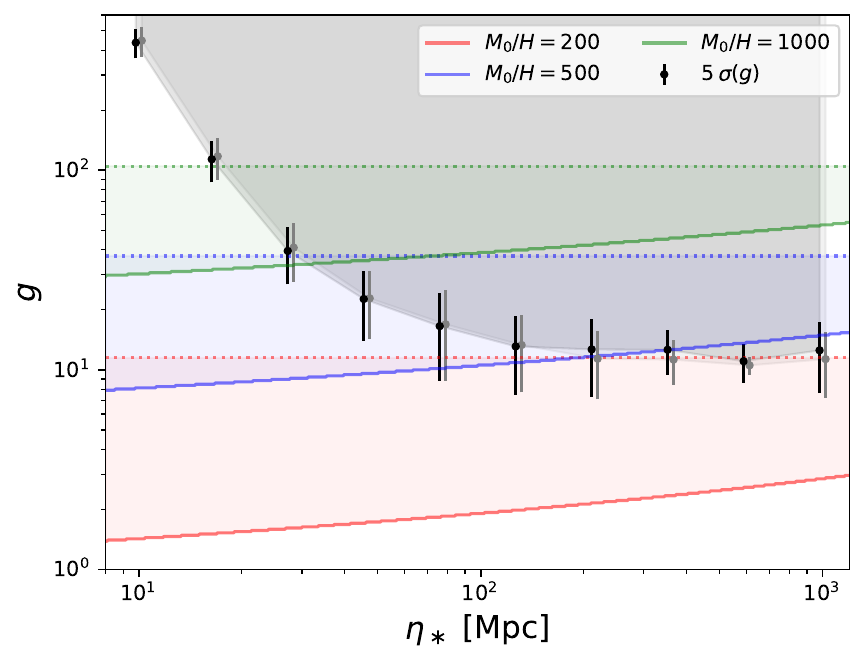}
    \includegraphics[width=0.49\textwidth]{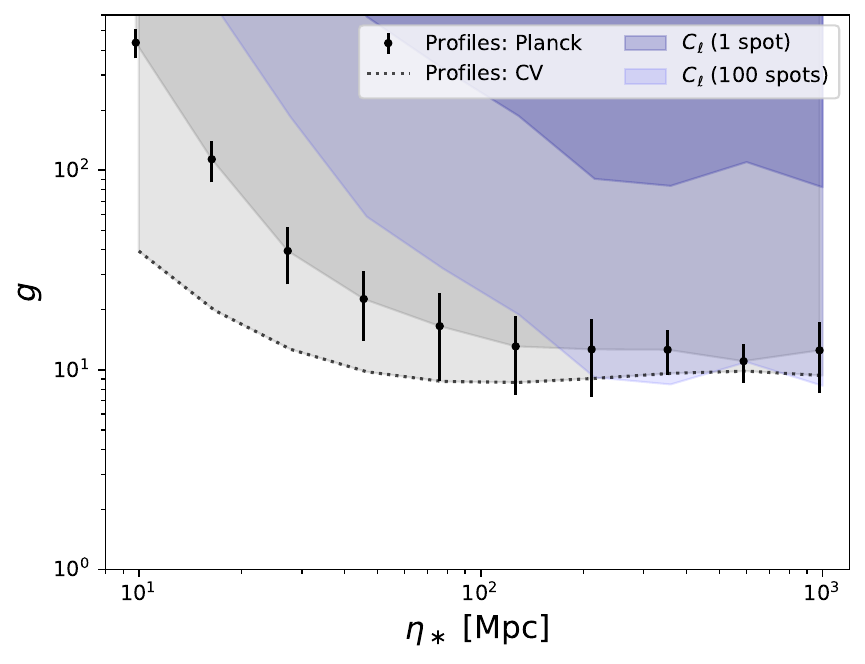}
    \caption{\textbf{Left panel}: Exclusion plot from \textit{Planck} (non-)detections. The black points and shaded region show the $5\sigma$ error bar on $g$ obtained from the matched-filter analysis at each value of $\eta_\star$ (with variations due to $\eta_{\rm HS}$ and foreground noise indicated by the error bars). If $g$ exceeds this bound, we would have detected any and all hotspots present in the unmasked sky. In gray, we show the same results but applied to the \textsc{sevem} component-separated data. The colored bands indicate the plausible regions of discovery space corresponding to different values of the heavy-particle mass $M_0$. The lower bound is set by requiring that at least one hotspot be produced (see Eq.~\eqref{eq: N-spots}), whilst the upper limit is set by backreaction constraints \citep{Kim:2021ida}. If the band lies above the \textit{Planck} limit, it can be robustly excluded. \textbf{Right panel}: forecasted exclusion from an idealized cosmic-variance-limited experiment for $\ell\in[2,3500]$ (dotted line) compared to the \textit{Planck} results. For small hotspots, we find significant improvements, though the largest hotspots are already close to cosmic-variance-limited. In blue, we show the $5\sigma$ bounds on $g$ possible from a power spectrum analysis; unlike for matched-filter analyses, these strengthen considerably if the number of hotspots (shown in the caption) is increased.}
    \label{fig: planck-exc}
\end{figure*}

Finally, we can consider the bounds on inflationary physics that can be extracted from these results. This is summarized in the left panel of Fig.~\ref{fig: planck-exc}, which shows the upper bounds on the coupling $g$ as a function of the hotspot size $\eta_\star$. These are obtained from the $\sigma(g)$ bounds computed in Eq.~\eqref{eq: matched-filter}, and show slight variations across the map, given the spatially-varying noise profile. As expected from the injection analysis, we find a bound of $g\lesssim 10$ for $\eta_\star\gtrsim 100\,\,\mathrm{Mpc}$ hotspots, though the constraint greatly weakens for small $\eta_\star$. We additionally find almost identical results from both the full-frequency and component-separated data, validating our approach. These upper bounds may be compared to the expected values of $g$, given that we wish to produce at least one hotspot (see Eq.~\eqref{eq: N-spots}) and must avoid backreaction of hotspots on the inflationary dynamics \citep{Kim:2021ida}. Whilst our bounds are unable to constrain models with $M_0/H_I\approx 200$, they effectively rule out hotspot production from models with $M_0/H_I\gtrsim500$ at $\eta_\star\gtrsim100\,\,\mathrm{Mpc}$. The ability to constrain \textit{extremely} massive particles arises since hotspot production in those models requires very high $g$.

\section{Summary \& Discussion}
\noindent The inflationary landscape is likely both non-trivial and non-linear. Whilst searches for low-order primordial non-Gaussianity place a number of restrictions on this space, many inflationary phenomena remain unconstrained. Here, we have considered one such example: non-adiabatic massive particle production from the inflationary vacuum (here realized via time-dependent masses, which break shift symmetries). If the particle is sufficiently massive, this generically leads to spatially-localized features in the distribution of primordial perturbations, which, unless subject to a dedicated search, can easily go unnoticed. Working in the context of \citep{Kim:2021ida, Kim:2023wuk}, we have performed the first searches for such models (in a manner analogous to \citep{Flauger:2016idt, Munchmeyer:2019wlh}, which focused on a model of periodic particle production and \emph{WMAP} data).
We find that even a single instance of particle production when CMB-observable modes exit the horizon can be effectively constrained using the {\it Planck} data.
In comparison to the earlier studies that consider particle production through vanishing particle masses during certain epochs, in the present scenario particle masses never become vanishingly small.
Correspondingly, the dominant signature is in the form of rare localized hot (or cold) spots, as opposed to having an observable feature in the curvature perturbation power spectrum.

Searching for localized features in the CMB is a task well-known to cluster cosmologists. Taking advantage of this parallel, we have utilized codes developed for thermal Sunyaev-Zel'dovich analysis to search for particle-production signatures in \textit{Planck} PR4 data using matched-filter methods. It is difficult to overemphasize the simplicity of this approach: one simply replaces a tSZ cluster profile (and tSZ frequency dependence) with the desired primordial template (and blackbody frequency dependence). Extensive verification has demonstrated that this approach works: we can recover the input parameters of inflationary hotspots (both singlets and pairs) injected in realistic simulations, given sufficiently large couplings. From the \textit{Planck} data, we find no evidence for new physics, with no robust detection of hotspots with $\mathrm{SNR}\geq 6$ (using a high threshold due to the four free parameters in the model). This yields a bound on shift-symmetry-violating couplings between the inflaton and a much heavier field, with $g\lesssim 10$ at time $\eta_\star>100\,\,\mathrm{Mpc}$.  Assuming $60$ $e$-foldings of inflation between the horizon exit of a mode with physical size $1/H_0$ (the present day inverse Hubble parameter) and the end of inflation, $\eta_\star=100$~Mpc corresponds to $\approx 56$ $e$-foldings before the end of inflation.

This rules out the presence of any hotspots from an extremely massive inflationary field ($M_0\gtrsim 500H_I$), again for $\eta_\star>100\,\,\mathrm{Mpc}$. As an illustration given the current upper limit $H_I < 4.8\times 10^{13}$~GeV on the scale of inflation~\cite{BICEP:2021xfz}, this implies a {\it direct} sensitivity to particles with $M_0 \sim 2 \times 10^{16}$~GeV, of order the Grand Unified Theory scale.

An important feature of the profile-finding analyses of this work is that their sensitivity does not depend on the number of hotspots present; we can detect \textit{any} hotspot given sufficiently large $g,\eta_{\star}$. This differs significantly from power-spectrum-based analyses, whereupon the modification to $C_\ell^{TT}$ scales as $g^2N_{\rm spots}$ \citep[{\it e.g.},][]{Kim:2021ida}. In the right panel of Fig.~\ref{fig: planck-exc}, we compare the $5\sigma$ bound on $g$ from the two approaches, considering both one and one hundred hotspots. In the former case, the constraints from the localized profile-finding analysis dwarf those from the power spectrum, whilst in the latter case, searching for modifications to $C_{\ell}^{TT}$ may yield improved bounds, particularly at high $\eta_\star$. This indicates an important and generic point: rare events are best searched for via local-in-space analyses, whilst common events (giving some spectrum of sources) can be probed best with low-order correlators.

The analysis developed herein can be extended in a number of ways. Firstly, we have considered only temperature anisotropies; the \textit{Planck} CMB dataset also contains polarization information, which can also be used to probe primordial hotspots since our mechanism produces adiabatic fluctuations that contribute to both $T$ and $E$. This will be particularly relevant for future surveys since large-scale temperature anisotropies are already cosmic-variance-limited. Furthermore, high-resolution data from the Atacama Cosmology Telescope~\cite{Henderson:2015nzj,ACT:2023wcq}, South Pole Telescope~\cite{SPT-3G:2014dbx,SPT-SZ:2021gsa}, Simons Observatory~\cite{SimonsObservatory:2018koc}, and CMB-S4~\cite{Abazajian:2019eic} will facilitate analysis of smaller hotspots, \resub{whose spatial signatures are} indistinguishable from point sources at the \emph{Planck} resolution. This is shown in the right panel of Fig.~\ref{fig: planck-exc}, which gives the upper bounds on $g$ from an idealized experiment with neither a beam nor experimental noise for $\ell\in[2,3500]$. As shown, one could improve the limit on $g$ for small $\eta_\star$ by at least an order of magnitude in the future, though cosmic variance limits improvements at large $\eta_\star$. Finally, we note that our pipeline can be simply extended to a wide variety of other primordial features, for example those of \citep{Flauger:2016idt,Munchmeyer:2019wlh}, allowing new insights into physics at the highest energies.

\vskip 16pt
\acknowledgments
{\footnotesize
\noindent 
We thank Taegyun Kim and Neal Weiner for insightful comments and discussions, \resub{as well as the anonymous referee for insightful feedback}. OHEP is a Junior Fellow of the Simons Society of Fellows and thanks Bandai Namco Entertainment Inc.~for inspiring the NYC particle astrophysics and cosmology group meeting, as well as Netflix for motivating the title. SK was supported in part by NSF grant PHY-2210498 and the Simons Foundation. JCH acknowledges support from NSF grant AST-2108536, DOE grant DE-SC0011941, NASA grants 21-ATP21-0129 and 22-ADAP22-0145, the Sloan Foundation, and the Simons Foundation. This work utilized {\tt numpy}~\cite{harris2020array}, {\tt matplotlib}~\cite{Hunter:2007}, {\tt healpy}~\cite{Zonca2019}, and {\tt HEALPix}~\cite{2005ApJ...622..759G}.}

\appendix
\section{Visual Inspection of \textit{Planck} Hotspot Candidates}
\begin{figure*}
    \centering
    \includegraphics[width=0.9\textwidth]{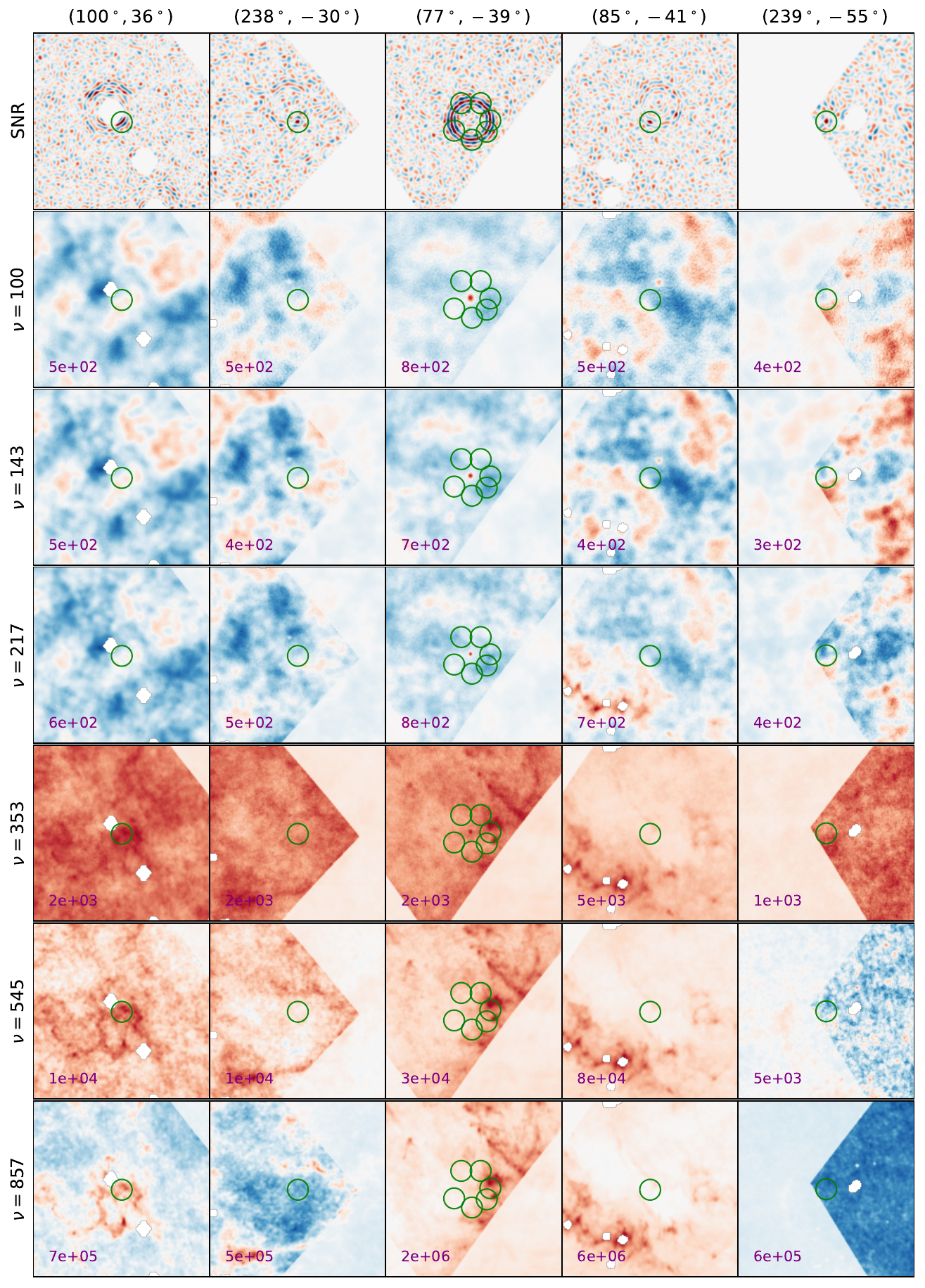}
    \caption{SNR maps (top row) and frequency maps (other rows, with frequency indicated in GHz) displaying the $\mathrm{SNR}
    \geq 6$ hotspot candidates, which are circled in green. Each image is a $5^\circ\times5^\circ$ Cartesian projection around the centers given in the title. The SNR maps have colorbar truncated to $\pm5$, whilst the maximal value for each other plot is shown in purple. The bounding rectangles demarcate the flat-sky tiles used in the analysis, and the holes are due to point source masking.}
    \label{fig: inspection}
\end{figure*}

\noindent In Fig.~\ref{fig: inspection}, we plot Cartesian projections of the \textit{Planck} HFI frequency maps centered on the $\mathrm{SNR}\geq 6$ hotspot candidates given in Tab.~\ref{tab: candidates}. We also show the SNR map (given by $\hat{g}/\sigma(g)$, as defined in Eq.~\eqref{eq: matched-filter} with optimal $\eta_\star,\eta_{\rm HS}$), computed for the $14.8^\circ \times 14.8^\circ$ tile in which the candidate was found. The first column appears consistent with a clump of dust (seen clearly at high frequency). Furthermore, we see a ringing feature for the third, which circles a synchrotron source identified in \citep{Planck:2018kuk}. For the other sources, visual inspection does not yield definitive results.

\bibliographystyle{apsrev4-1}
\bibliography{refs}

\end{document}